\documentclass[12pt]{article}

\usepackage{amsmath,amsfonts,amssymb,latexsym}

\def\be{\begin{equation}}
\def\ee{\end{equation}}
\def\bq{\begin{eqnarray}}
\def\eq{\end{eqnarray}}
\def\beq{\begin{eqnarray*}}
\def\eeq{\end{eqnarray*}}
\newcommand{\GA}{\alpha}
\newcommand{\GB}{\beta}
\newcommand{\GG}{\gamma}
\newcommand{\GD}{\delta}

\newcommand{\GK}{\kappa}

\newcommand{\GR}{\rho}

\newcommand{\GT}{\tau}

\begin{document}



\begin{center}
{\huge Asymptotics of flat, radiation universes in quadratic
gravity}

\vspace{1cm}

{\large Spiros Cotsakis$\dagger$ and Antonios Tsokaros$\ddagger$}\\

\vspace{0.5cm}

{\normalsize {\em Research Group of Geometry, Dynamical Systems and
Cosmology}}\\ {\normalsize {\em Department of Information and
Communication Systems Engineering}}\\ {\normalsize {\em University
of the Aegean}}\\ {\normalsize {\em Karlovassi 83 200, Samos,
Greece}}\\ {\normalsize {\em E-mails:} $\dagger$
\texttt{skot@aegean.gr}, \texttt{$\ddagger$ atsok@aegean.gr}}
\end{center}

\begin{abstract}
\noindent We consider the asymptotics of flat, radiation-dominated
isotropic universes in four-dimensional theories with quadratic
curvature corrections which may arise when contributions related to
the string parameter $\alpha'$  are switched on. We show that all
such universes are singular initially, and calculate all  early time
asymptotics using the method of asymptotic splittings. We also
examine the late time asymptotic behaviour  of these models and show
that there are no solutions which diverge as $e^{t^2}$, the known
radiation solution of general relativity is essentially the only
late time asymptotic possibility in these models.
\end{abstract}


\section{Introduction}
\label{sec:intro} A marked feature of theoretical cosmology during
the last few decades has been the proliferation of different
dynamical theories of gravity that are used to tackle the basic
cosmological issues. With the advent of string theory and its
subsequent generalizations, these theories acquired a status similar
to the shafts and knobs of  J. L. Synge's allegorical box, an
unopened (perhaps unopenable) box in which a vast number of shafts
are connected to one another by many complicated laws, making it
objectively impossible to explain what really is inside. There is,
it is true, an appealing commonsense in this new picture and,
although the promise of hard rationality is not yet fulfilled, one
hopes that in time some new forms will develop and the whole picture
will come into better focus.

Another feature common to many string theory effective actions  is
the presence of higher order curvature corrections, the importance
of which cannot be neglected unless the basic  string parameter
$\alpha'$ is small. The generic form of the higher order curvature terms
present in these effective string actions  is that of a typical
Gauss-Bonnet combination, cf. \cite{1,2}. Such theories admit
constant curvature vacua even in the absence of a cosmological
constant, many of which are obviously stable \cite{deser}.

The pertinent issue of the structure and nature of cosmological
singularities becomes of interest in this new context for at least
two reasons. First, it is important to know how different asymptotic
behaviours can be compared to the known situation in general
relativity, for instance whether or not simple isotropic solutions
act as past attractors to more general homogeneous universes in
quadratic theory, cf. \cite{fle1,fle2,bh}. Secondly, in the context
of string theory the basic problem becomes that of understanding the
structure of cosmological singularities and how the latter could be
resolved using the notion of string dualities, cf. \cite{w}.

In this note we face the more humble task of analyzing the
asymptotic form of certain simple cosmological solutions to the
general quadratic theory in four dimensions. Our analysis includes
both late and early time asymptotics. We shall show that contrary to
claims usually made in the literature of such models (such claims
were first made as early as 1969, cf. \cite{RR}), one cannot
conclude that solutions (which are regular initially) fail to
approach the general relativistic solution at late times. In fact,
we show that there can be no regular solutions at early times
and by applying the method of asymptotic splittings  of \cite{CB},
we further prove that there is an open set of initial conditions for
which the general solution blows up at the big bang (collapse)
singularity.

The plan of this short paper is as follows. In the next Section we
write down the basic equations of the quadratic theory which we
consider further in later Sections. We then show that their late
time asymptotics cannot be of the divergent form $e^{t^ 2}$,
contrary to known claims found in the literature. In Section 3, we
apply the method of asymptotic splittings to find the form of the
general solution of these models near the collapse singularity in
the past. We may therefore conclude that simple bouncing, flat,
radiation universes with quadratic curvature corrections do not
exist.

\section{Field equations and late time asymptotics}
We start with the action \be\mathcal{S}=\int_{\mathcal{M}^4}\mathcal{L}_{\textrm{total}}d\mu_{g},\quad \quad d\mu_{g}=\sqrt{-g} d\Omega,\ee
where $\mathcal{L}_{\textrm{total}}$ is the lagrangian density of the general
quadratic gravity theory given in the form\footnote{the conventions
for the metric and the Riemann tensor are those of \cite{LL}.} $\mathcal{L}_{\textrm{total}}=\mathcal{L}(R)+\mathcal{L}_{\textrm{matter}}$, with \be
\mathcal{L}(R)=R + BR^2 + C\textrm{Ric}^2 + D\textrm{Riem}^2 ,
\label{eq:lagra} \ee
where $B,C,D$ are constants. Since in four dimensions we have the Gauss-Bonnet  identity,
\be \GD \int_{\mathcal{M}^4}R^2_{GB}d\mu_{g}=0,\quad
R^2_{GB}=R^2 - 4\textrm{Ric}^2 + \textrm{Riem}^2,
\label{eq:gentity} \ee
in the derivation of the field equations
through variation  of the action associated with (\ref{eq:lagra}),
only terms up to $\textrm{Ric}^2$ will matter. Therefore the variational derivative of the action leads to the following field equations:
\begin{eqnarray}
&&\frac{8\pi G}{c^4}T^{ij}=R^{ij}-\frac{1}{2}g^{ij}R  +B \left[2RR^{ij}-\frac{1}{2}R^2g^{ij}-2(g^{ik}g^{jm}-g^{ij}g^{km})\nabla_{k}\nabla_{m}R \right] \nonumber \\
    &+& C \left[2R^{ik}R^{j}_{\: \: k}-\frac{1}{2}g^{ij}R^{km}R_{km}+\nabla_k\nabla^k R^{ij}
              + g^{ij}\nabla_k\nabla_m R^{mk} - 2\nabla_k \nabla^j R^{ki}  \right]. \label{eq:fe1} \end{eqnarray}
Below we focus exclusively in spatially flat universes of the
form
\be ds^2=dt^2-b(t)^2(dx^2+dy^2+dz^2), \label{eq:flatun}\ee which
are radiation dominated ($P=\GR/3$). For such spacetimes
 we have a second useful identity,
 \be \GD \int_{\mathcal{M}^4} (R^2 - 3\textrm{Ric}^2)d\mu_{g}=0 \: ,
\label{eq:isontity} \ee which further enables us to include the
contribution of the $\textrm{Ric}^2$ term into the coefficient of
$R^2$, altering only the arbitrary constants. In this case the field
equations (\ref{eq:fe1}) simplify as follows: \be
R^{ij}-\frac{1}{2}g^{ij}R+
      \frac{\GK}{6} \left[2RR^{ij}-\frac{1}{2}R^2g^{ij}-2(g^{ik}g^{jm}-g^{ij}g^{km})\nabla_{k}\nabla_{m}R \right]=\frac{8\pi G}{c^4}T^{ij},
\label{eq:fe} \ee where $\GK=6B+2C$. This naturally splits into
$00-$ and $ii-$components, but only the $00-$component of
(\ref{eq:fe}) will be used below. This reads: \be
\frac{\dot{b}^2}{b^2}-\GK\left[2\: \frac{\dddot{b}\:\dot{b}}{b^2} +
2\:\frac{\ddot{b}\dot{b}^2}{b^3}-\frac{\ddot{b}^2}{b^2} - 3\:
\frac{\dot{b}^4}{b^4} \right] - \frac{b_{1}^2}{b^4} =0 ,
\label{eq:beq} \ee where $b_{1}$ is a constant defined by
\be\label{eq.9} \frac{8\pi G \rho}{3c^4}=\frac{b_{1}^2}{b^4},\quad
(\textrm{from}\,\,\nabla_{i}T^{i0}=0). \ee Note that the Friedmann
solution $\sqrt{2b_1 t}$ of general relativity satisfies the above
equation.

In the rest of this Section we shall consider the problem of the
late time asymptotics of solutions to Eq. (\ref{eq:beq}). We say
that a solution $b(t)$ is \emph{asymptotic} to another solution
$a(t)$ provided that the following two conditions hold (the first is
subdivided):
\begin{enumerate}
\item[(i)] Either $(1)$ $a(t)$ is an exact solution of the system, or
 $(2)$ $a(t)$ is a solution of the system (substitution gives $0=0$) as $t\rightarrow\infty$,
\item [(ii)] $b(t)=a(t)[1+g(t)],\, g(t)\rightarrow 0$, as $t$ tends to infinity.
\end{enumerate}
If either of these two conditions is not satisfied, then $b(t)$
cannot be asymptotic to $a(t)$.

Let us assume, following \cite{RR},  that Eq. (\ref{eq:beq}) has a solution with a regular
minimum at the arbitrary time $t_0$, ($\dot{b}_0\equiv\dot{b}(t=t_0)=0$ and
$b_0\equiv b(t=t_0)\neq 0$).  We can then expand this solution as a Taylor series
\be b(t)=b_{0} + \frac{\ddot{b}_{0}}{2}\: (t-t_0)^{2} +
\frac{\dddot{b}_{0}}{6}\: (t-t_0)^{3} + \cdots \:\:,
\label{eq:taylor} \ee and substitute this form back to Eq. (\ref{eq:beq}), to see that this
restricts the value of the constant $\GK$ to
$\GK=(b_{1}/(b_{0}\ddot{b}_{0}))^{2} > 0$.

Reduction of the order of Eq. (\ref{eq:beq}) can be achieved if we
set $f=(b\dot{b})^{3/2}$ and $\xi=12^{-3/4}b^3$, so as to obtain the
following second order differential equation: \be
f''-\frac{1}{\GK\xi^{2/3}}(f^{-1/3}-b_{1}^{2}f^{-5/3})=0\: ,
\label{eq:feq} \ee where the derivative is with respect to $\xi$.
For large values of $\xi$, an exact asymptotic solution of
(\ref{eq:feq}), in the sense of condition $(\textrm{i}2)$ above,  is
given by \be f \sim \left(\frac{4}{3\GK}\right)^{3/4} \xi
(\ln\xi)^{3/4}, \quad\quad \textrm{for large } \xi, \label{eq:fas}
\ee which in terms of the variables $b$ and $t$ has the form \be
b(t) \sim 12^{1/4} e^{(t-t_0)^2/12\GK}. \label{eq:bas} \ee However,
we stress that \textit{this is no more an asymptotic solution of the
original equation} (\ref{eq:beq}) (as it is, for instance, claimed
in \cite{RR}). In fact, a simple substitution of (\ref{eq:bas}) into
the left hand side of Eq. (\ref{eq:beq}) gives the result:
$$-b_{1}^{2} e^{-(t-t_0)^2/3\GK}+\frac{1}{36\GK},$$ which never
goes over to zero as $t$ tends to infinity, so that condition
$(\textrm{i}2)$ can never be satisfied. In other words, an
asymptotic solution of Eq. (\ref{eq:feq}) does not necessarily
translate into an asymptotic solution of Eq. (\ref{eq:beq}), and
this is the case for (\ref{eq:fas}). These  results lead us to
conclude that there exist no solutions of the field equation
(\ref{eq:feq}) which have late time asymptotics of the form
(\ref{eq:bas}).

One candidate solution (leading to a well-posed asymptotic problem
for late or early times) that satisfies condition $(\textrm{i})$ of
the definition and so qualifies to perturb is the radiation solution
proportional to $\sqrt{t}$ as we already noted  after Eq.
(\ref{eq.9}). The problem of the late (as well as early) time
asymptotics of solutions to the field equations of the quadratic
theories
 was taken up in Refs. \cite{fle1, fle2} using a detailed
perturbation analysis of the FRW radiation solutions (flat and
non-flat) of the form $a(t)=(t-\sigma t^2)^{1/2}$ (in our present
work we focus in the flat case, $\sigma =0$). The conclusion is
that, although the non-flat radiation solutions are generically
unstable, in the case of a flat universe there is a parameter region
in which all late time solutions are stable with respect to
perturbations, the latter decaying as $g(t)\sim t^{-1} +t^{-3/4}$,
(cf., e.g., \cite{fle2}, Eq. (21)). Therefore in such models all
relevant solutions of our higher order gravity theory asymptotically
approach the flat, radiation solution of general relativity. In view
of the non-existence of an $e^{t^2}$ late time asymptotic shown
currently, these perturbation conclusions suggest that  the flat,
radiation form $\sqrt{t}$ is a unique late time asymptotic solution
in the category considered here.

\section{Early  asymptotic splittings}\label{sec:earlyTA}
We now move on to  perform a local asymptotic analysis in
order to find the general behaviour of the solutions of Eq.
(\ref{eq:beq}) near the initial singularity. This analysis is based
on the use of the method of asymptotic splittings expounded in Ref.
\cite{CB}, we follow their notation closely. As a first step, after setting $b=x$, $\dot{b}=y$ and
$\ddot{b}=z$, Eq. (\ref{eq:beq}) can be written as a dynamical
system of the form $\mathbf{\dot{x}}=\mathbf{f}(\mathbf{x})$, $\mathbf{x}=(x,y,z)$:
\begin{equation}
\label{eq:ds}
\dot{x} = y,\:\:\:\:\: \dot{y} = z,\:\:\:\:\: \dot{z} = \frac{y}{2\GK} - \frac{b_{1}^2}{2\GK yx^2} - \frac{yz}{x} + \frac{z^2}{2y} + \frac{3y^3}{2x^2}.
\end{equation}
If $\mathbf{a}=(\GA, \GB, \GG)$, and $\mathbf{p}=(p, q, r)$, we
denote by $\mathbf{x}(\GT)$ the solution \be
\mathbf{x}(\GT)=\mathbf{a}\GT^{\mathbf{p}}=(\GA \GT^{p}, \GB
\GT^{q}, \GG \GT^{r}), \label{eq:domisol} \ee and by direct
substitution in our system (\ref{eq:ds}), we look for the possible
scale invariant solutions of this form\footnote{The vector field
$\mathbf{f}$ is called scale invariant if
$\mathbf{f}(\mathbf{a}\GT^{\mathbf{p}})=\GT^{\mathbf{p-1}}
\mathbf{f}(\mathbf{a})$}. There are two possible combinations. The
first, which is the most interesting one, has \emph{dominant part}
\be
\mathbf{f}^{(0)}=\left(y,z,\frac{z^2}{2y}-\frac{zy}{x}+\frac{3y^3}{2x^2}
\right), \label{eq:f0} \ee while the subdominant part reads \be
\mathbf{f}^{\,\textrm{sub}}=\left(0,0,-\frac{b_{1}^{2}}{2\GK
yx^2}+\frac{y}{2\GK}  \right), \ee with
$\mathbf{f}=\mathbf{f}^{(0)} + \mathbf{f}^{\,\textrm{sub}}$. The
dominant balance (of order 3) turns out to be \be
(\mathbf{a},\mathbf{p}) = \left( \left(
\GA,\frac{\GA}{2},-\frac{\GA}{4}\right),\:
\left(\frac{1}{2},-\frac{1}{2},-\frac{3}{2} \right) \right),
\label{eq:domibal} \ee where $\GA$ is an arbitrary constant. We
recognize this as being what we want, but we are not done yet. We now
calculate the eigenvalues of an important matrix, called the
K-matrix, which signify the places in a series expansion of the
solution around the finite time singularity where arbitrary
constants appear, offering thus a clue as to how general the found
solution is. The Kowalevskaya exponents for this particular
decomposition, eigenvalues of the matrix
$\mathcal{K}=D\mathbf{f}(\mathbf{a})-\textrm{diag}(\mathbf{p})$, are
$\{-1,0,3/2\}$ with corresponding eigenvectors
$\{(4,-2,3),(4,2,-1),(1,2,2)\}$. The arbitrariness coming from the
coefficient $\GA$ in the dominant balance reflects the fact that one
of the dominant exponents is zero with multiplicity one.

Keeping with the method of asymptotic splittings \cite{CB}, we
proceed to construct series expansions which are local solutions
around movable singularities. In our particular problem, the
expansion around the singularity turns out to be a \emph{Puiseux series}
 of the form
\begin{equation} \label{eq:series}
x(t) = \sum_{i=0}^{\infty} c_{1i} (t-t_{0})^{\frac{i}{2}+\frac{1}{2}}, \:\:\:\:\:
y(t) = \sum_{i=0}^{\infty} c_{2i}  (t-t_{0})^{\frac{i}{2}-\frac{1}{2}},\:\:\:\:\:
z(t) = \sum_{i=0}^{\infty} c_{3i} (t-t_{0})^{\frac{i}{2}-\frac{3}{2}} ,\:\:\:\:\:
\end{equation}
where $t_0$ is arbitrary and $c_{10}=\GA ,c_{20}=\GA /2,c_{30}=-\GA
/4$. For these series expansions to be valid \emph{the compatibility
condition} \be (1,2,2) \cdot \left( \begin{array}{l}
                       -2c_{13}+c_{23}  \\
                       -c_{23}+c_{33}  \\
                       -\frac{1}{2}c_{13}+\frac{5}{4}c_{23}-c_{33}
                     \end{array}
              \right) = 0,
\label{eq:cc} \ee must be satisfied. Substitution of Eq.
(\ref{eq:series}) into Eq. (\ref{eq:ds}) leads  to recursion
relations that determine the unknowns $c_{1i}, c_{2i}, c_{3i}$ term
by term. After verifying that Eq. (\ref{eq:cc}) is indeed true, we may
write the final series expansion corresponding to the balance
(\ref{eq:domibal}). It is: \be x(t) = \GA \:\:
(t-t_{0})^{\frac{1}{2}} + c_{13} \:\: (t-t_{0})^{2} + \displaystyle
\frac{\GA^4-4b_{1}^2}{24\GK\GA^3} \:\:
                    (t-t_{0})^{\frac{5}{2}} + \cdots .
\label{eq:sol} \ee The series expansions for $y(t)$ and $z(t)$ are
given by  the first and second time derivatives of the above
expressions respectively.

Our series (\ref{eq:sol}) has three arbitrary constants, $\GA,
c_{13}, t_0$ (the last corresponding to the arbitrary position of
the singularity) and is therefore a local expansion of the
\emph{general} solution around the movable singularity $t_0$. Since
the leading order coefficients can be taken to be real, by a theorem
of Goriely-Hyde \cite{GH}, we conclude that there is an open set of
 initial conditions  for which the general solution blows up at the
finite time (initial) singularity at $t_0$. Finally, we observe that
near the initial singularity, all flat, radiation solutions of the
quadratic gravity theory considered here are Friedmann-like
regardless of the sign of the $R^2$ coefficient, while away from the
singularity they strongly diverge from such forms. This proves the
stability of our solutions in the neighborhood of the singularity.

The second possibility of a scale invariant solution has
dominant part \be
\mathbf{f}^{(0)}=\left(y,z,-\frac{b_{1}^{2}}{2\GK
yx^2}+\frac{z^2}{2y}-\frac{zy}{x}+\frac{3y^3}{2x^2}  \right), \ee and
subdominant part \be
\mathbf{f}^{\,\textrm{sub}}=\left(0,0,\frac{y}{2\GK}  \right). \ee
The dominant balance (of order 2)  at the singularity turns out to be
\be \{
(\mathbf{a},\mathbf{p}) \} = \{ ( (\GA,\GA,0),\: (1,0,-1) ), (
(-\GA,-\GA,0),\: (1,0,-1) ) \}, \ee where $\GA$ a constant defined by
$\GA=(b_{1}^2/3\GK)^{1/4}\in \mathbb{R}$. The Kowalevskaya
exponents are $\{-1,\sqrt{6},-\sqrt{6}\}$, therefore this
decomposition leads to a \emph{particular} solution (two arbitrary
constants) whose dominant term near the initial singularity is
$b(t)\sim (t-t_0)$ and therefore does not provide any bounce
solutions.

\section{Conclusions}
We have considered the behaviour of flat, radiation-dominated
solutions to the general quadratic theory of gravity in four
dimensions. This theory is supposed to arise as part of  an
effective string theory action truncated to first order in
$\alpha'$. We found that in these models there are no bounce
solutions, all have a collapse, initial, isotropic singularity and
we have given explicit forms of the approach to the initial finite
time singularity. However, our results suggest that it is not
possible to inhibit the existence of late time behaviour similar to
the observed, even though any such solution must necessarily be
singular initially.

There are many ways to extend these results. An obvious one is to
take into account the dilaton dependence present in the full string
action (cf. \cite{1}, Eq. (4.3)), in other words to consider the
present problem in the string frame (and not in the Einstein
conformal frame of the present paper). We also wish to see how our
results are altered (or not altered) when we pass on to models with
curvature, especially near the singularity.

\section*{Acknowledgments}
\noindent The authors thank P. G. L. Leach, J. Miritzis, D. F. Mota
and R. Olea for useful discussions and correspondence. We also thank
an anonymous referee whose comments led to an improved version of
Section 2.  Part of the work of S.C. was done while visiting the
IHES in Bures sur Yvette. This work was co-funded by 75\% from the
EU and 25\% from the Greek Government, under the framework of the
``EPEAEK: Education and initial vocational training program -
Pythagoras''.

\end{document}